\documentclass[onecolumn,prd,nofootinbib,showpacs,floatfix]{revtex4}
\usepackage{epsfig}
\newcommand{\be}{\begin{equation}}
\newcommand{\ee}{\end{equation}}
\newcommand{\bea}{\begin{eqnarray}}
\newcommand{\eea}{\end{eqnarray}}
\newcommand{\eqn}[1]{(\ref{#1})}
\newcommand{\pth}{\texttt{PArthENoPE} }
\newcommand{\dneff}{\Delta N_{\rm eff}}
\newcommand{\pp}{~~~.}
\newcommand{\vv}{~~~,}
\newcommand{\p}{{\rm p}}
\newcommand{\lrt}{\leftrightarrow}
\newcommand{\rt}{\rightarrow}
\newcommand{\pe}{{\phi_e}}
\newcommand{\hrho}{\hat{\rho}}
\newcommand{\rhol}{\rho_\Lambda}
\newcommand{\lh}{\hat{L}}
\newcommand{\ds}{\displaystyle}
\newcommand{\hH}{\widehat{H}}
\newcommand{\hp}{\hat{\rm p}}
\newcommand{\hmu}{\widehat{M}_u}
\newcommand{\hdmi}{\Delta \widehat{M}_i}

\newcommand{\hgi}{\widehat{\Gamma}_i}

\newcommand{\hnb}{\hat{n}_B}
\newcommand{\nn}{\nonumber}

\newcommand{\etal}{{\it et al.}}
\newcommand{\ApJ}{{ Astrophys. J.\,}}
\newcommand{\ApJS}{{Astrophys. J. Suppl.\,}}

\newcommand{\NP}{{ Nucl. Phys.\,}}
\newcommand{\PR}{{Phys. Rev.\,}}

\newcommand{\PL}{{Phys. Lett.\,}}

\begin{document}

\title{\pth: \texttt{P}ublic \texttt{A}lgo\texttt{r}i\texttt{th}m \texttt{E}valuating the \texttt{N}ucleosynthesis \texttt{o}f
\texttt{P}rimordial \texttt{E}lements}

\author{O. Pisanti$^{1}$\footnote{Corresponding author. E-mail: pisanti@na.infn.it},
A. Cirillo$^{1}$, S. Esposito$^{1}$, F. Iocco$^{1,2}$, G. Mangano$^{1}$, G. Miele$^{1}$,
and P.~D.~Serpico$^{3}$}
\affiliation{ $^1$Dipartimento di Scienze Fisiche, Universit\`a di Napoli Federico II\\
and INFN, Sezione di Napoli, Via Cintia, I-80126 Napoli, Italy\\
$^2$Kavli Institute for Particle Astrophysics and Cosmology,
PO Box 20450, Stanford, CA 94309, USA\\
$^3$Center for Particle Astrophysics, Fermi National Accelerator
Laboratory, Batavia, IL 60510-0500, USA}

\begin{abstract}
We describe a program for computing the abundances of light
elements produced during Big Bang Nucleosynthesis which is
publicly available at http://parthenope.na.infn.it/. Starting from
nuclear statistical equilibrium conditions the program solves the
set of coupled ordinary differential equations, follows the
departure from chemical equilibrium of nuclear species, and
determines their asymptotic abundances as function of several
input cosmological parameters as the baryon density, the number of
effective neutrino, the value of cosmological constant and the
neutrino chemical potential. The program requires commercial NAG
library routines.

\section*{Program summary}
\noindent
{\it Title of program}:
\pth \\
{\it Program URL}: http://parthenope.na.infn.it/\\
{\it Program obtainable from}: parthenope@na.infn.it\\
{\it Computers}: PC-compatible running Fortran on Unix, MS Windows or Linux\\
{\it Operating systems under which the program has been tested:} Windows 2000, Windows XP, Linux\\
{\it Programming language used}: Fortran 77\\
{\it External routines/libraries used}: NAG libraries\\
{\it No. of lines in distributed program, including input card and test data}: 4969\\
{\it No. of bytes in distributed program, including input card and test data}: 192 Kb\\
{\it Distribution format}: tar.gz\\
{\it Nature of physical problem}: Computation of yields of light elements synthesized in the primordial universe.\\
{\it Method of solution:} BDF method for the integration of the ODE's, implemented in a NAG routine\\
{\it Typical running time:} 90 sec with default parameters on a Dual Xeon Processor 2.4GHz with 2.GB RAM

\end{abstract}

\pacs{26.35.+c %Big Bang nucleosynthesis
 \hfill DSF 13/07, FERMILAB-PUB-07-079-A, SLAC-PUB-12488}

\maketitle
\begin{quote}
\begin{flushleft}
{\it ``A l'alta fantasia qui manc\`o possa;\\
ma gi\`a volgeva il mio disio e 'l velle,\\
s\`i come rota ch'igualmente \`e mossa,\\
l'amor che move il sole e l' altre stelle."}

Dante Alighieri, ``Commedia" - Paradiso, Canto XXXIII, 142-145
\end{flushleft}
\end{quote}

\section{Introduction}
\label{intr}

Big Bang Nucleosynthesis (BBN) is one of the fundamental pillars
of the Cosmological Standard Model. In the very early Universe,
when the temperature of the primordial plasma decreased from a few
MeV down to $\sim 10\,$keV, light nuclides as $^2$H, $^3$He,
$^4$He and, to a smaller extent, $^7$Li were produced via a
network of nuclear processes. The relative abundances of these
nuclear ``ashes'' with respect to hydrogen can be determined via
several observational techniques and in different astrophysical
environments. In the standard cosmological scenario and in the
framework of the electroweak Standard Model, the dynamics of this
phase is controlled by only one free parameter, the baryon to
photon number density, which can thus be fixed by fitting
experimental observations. This parameter can be also
independently measured with very high precision by Cosmic
Microwave Background anisotropies \cite{Spergel:2006hy} and the
agreement with BBN result is quite remarkable. For reviews see
e.g. \cite{Fields:2006ga} in \cite{Yao:2006px} or
\cite{Steigman:2005uz}.

More in general, the quality and quantity of new cosmological and
astrophysical data available in the last decade has led to an
overall consistent picture of the evolution of the Universe,
usually referred to as the ``concordance" model. This is based on
standard physics plus a few phenomenological parameters, for which
an underlying theory is however still missing. At present, one
thus faces the intriguing possibility that one might test models
which go beyond our present understanding of fundamental
interactions, in a way which is complementary to traditional
earth-based laboratory and accelerator approaches. This is
illustrated e.g. by the search for new light degrees of freedom
which might contribute to the total energy density in the Universe
in addition to photons and neutrinos. To pursue this programme, it
is crucial to achieve a high level of accuracy in theoretical
predictions for cosmological observables, at least at the level of
experimental uncertainties. In the case of BBN, many steps have
been done in this direction by a careful analysis of several key
aspects of the physics involved in the phenomenon. The accuracy of
the weak reactions which enter the neutron/proton chemical
equilibrium has been pushed well below the percent level
\cite{Lopez,EMMP1,Esposito:1999sz,Esposito:2000hh,Serpico:2004gx}.
Similarly, the neutrino decoupling has been carefully studied by
several authors by explicitly solving the corresponding kinetic
equations
\cite{Gnedin,Dolgov:1998,dolgovrep,Mangano:2001iu,Mangano:2005cc,Mangano:2006ar}.
These two issues are mainly affecting the prediction of $^4$He
mass fraction, which presently has a very small uncertainty, of
the order of 0.1 \%, due to the experimental uncertainty on
neutron lifetime. Finally, much study has been devoted to the
analysis of several nuclear reaction rates entering the BBN
network, as well as the corresponding uncertainties. This task
involves a careful study of the available data or predictions on
each reaction, including an update in light of new relevant
experimental measurements, the choice of a reasonable protocol to
combine them in order to obtain a best estimate and an error and,
finally, the calculation of the corresponding thermal averaged
rates. This issue has been extensively discussed in
\cite{Serpico:2004gx}, whose results have been used in the program
described in the present paper, and
\cite{Cyburt:2004cq,Coc:2003ce}. An important benchmark in this
development has been represented by the compilation of the NACRE
Collaboration database \cite{nacre}.

In view of all these recent developments, we believe that the
scientific community interested in BBN, in itself or as a tool to
constrain new physics beyond the Standard Model, might find useful
a new public BBN code which updates the pioneering achievements of
\cite{Wagoner,KawCode92,Smith:1992yy}\footnote{It is a pleasure to acknowledge
the public code of \cite{KawCode92} as the starting point for many
scholars interested in BBN, including the authors of the present
paper.}. For this reason we have publicly released a code we have
developed and continuously updated over almost a decade, which we
named \pth and can be obtained at the URL
http://parthenope.na.infn.it/. The aim of the present paper is to
give a general description of the program and how to use it. After
briefly summarizing in Section \ref{s:generalities} the
theoretical framework of BBN and all major improvements
implemented in \pth, we discuss in Section \ref{ss:nonstandard} a
few extensions of the minimal standard BBN scenario which are also
included in the code. In Section \ref{s:code} the main structures
of \pth are outlined, while a comparison with the public code of
\cite{KawCode92} is discussed in Section \ref{s:compare}. Finally,
in Section \ref{s:concl} we report our conclusions. Hereafter we
use natural units where the reduced Planck constant, the speed of
light and the Boltzmann constant are fixed to 1, i.e.
$\hbar=c=k_B=1$.

%%%%%%%%%%%%%%%%%%%%%%%%%%%%%%%%%%%%%%%%%%%%%%%%%%
\section{The theory of Big Bang Nucleosynthesis}
\label{s:generalities}
%%%%%%%%%%%%%%%%%%%%%%%%%%%%%%%%%%%%%%%%%%%%%%%%%%

\subsection{The set of equations}
\label{ss:equations}

We consider $N_{nuc}$ species of nuclides, whose number densities,
$n_i$, are normalized with respect to the total number density of
baryons $n_B$,
\be
X_i=\frac{n_i}{n_B} \quad\quad\quad i=n,\,p,\,^2{\rm H}, \, ...
\pp
\ee
The list of all nuclides which are typically included in BBN
analyses and considered in \pth is reported in Table
\ref{t:nuclnumb}.

In the (photon) temperature range of interest for BBN, $10 \,{\rm
MeV}> T > 0.01\, {\rm MeV}$, electrons and positrons are kept in
thermodynamical equilibrium with photons by fast electromagnetic
interactions and distributed according to a Fermi-Dirac
distribution function $f_{e^{\pm}}$, with chemical potential
$\mu_e$, parameterized in the following by the function $\phi_e
\equiv \mu_e/T$. The pressure and energy density of the
electromagnetic plasma ($e^{\pm}$ and $\gamma$) is calculated in
\pth by including the effect of finite temperature QED corrections
\cite{Mangano:2001iu}. Furthermore, electromagnetic and nuclear
scatterings keep the non-relativistic baryons in kinetic
equilibrium, and their energy density $\rho_B$ and pressure $\p_B$
are given by
\bea
\rho_B & = & \left[M_u + \sum_i \left( \Delta M_i + \frac{3}{2} \,
T \right)~ X_i \right] n_B\,\,\, , \label{neupress}\\
\p_B & = & T \, n_B \, \sum_i X_i \,\,\,, \label{barpress}
\eea
with $\Delta M_i$ and $M_u$ the i-th nuclide mass excess and the
atomic mass unit, respectively.

\begin{table*}[t]
\begin{center}
\begin{tabular}{lclclclclc}
\hline \boldmath{No.} & ~~\boldmath{Nuclide}~~ & \boldmath{No.} & ~~\boldmath{Nuclide}~~ & \boldmath{No.}
& ~~\boldmath{Nuclide}~~ & \boldmath{No.} & ~~\boldmath{Nuclide}~~ & \boldmath{No.} & ~~\boldmath{Nuclide}~~ \\
\hline 1 &n   &   7 & $^6$Li  &  13 & $^{10}$B  &  19 & $^{13}$C  &  25 & $^{15}$O \\
\hline 2 &p   &   8 & $^7$Li  &  14 & $^{11}$B  &  20 & $^{13}$N  &  26 & $^{16}$O \\
\hline 3 &$^2$H  &   9 & $^7$Be  &  15 & $^{11}$C  &  21 & $^{14}$C  &   &     \\
\hline 4 &$^3$H  &  10 & $^8$Li  &  16 & $^{12}$B  &  22 & $^{14}$N  &   &       \\
\hline 5 &$^3$He &  11 & $^8$B   &  17 & $^{12}$C  &  23 & $^{14}$O  &   &     \\
\hline 6 &$^4$He &  12 & $^9$Be  &  18 & $^{12}$N  &  24 & $^{15}$N  &   & \\
\hline
\end{tabular}
\end{center}
\caption{Nuclides considered in \pth.} \label{t:nuclnumb}
\end{table*}

The set of differential equations ruling primordial
nucleosynthesis is the following (see for example
\cite{Wagoner,Esposito:1999sz,Esposito:2000hh}):
\bea
&&\frac{\dot{a}}{a}  = H = \sqrt{\frac{8\, \pi G_N}{3}~ \rho} \vv
\label{e:drdt} \\
&&\frac{\dot{n}_B}{n_B} = -\, 3\, H \vv
\label{e:dnbdt} \\
&&\dot{\rho} = -\, 3 \, H~ (\rho + \p) \vv
\label{e:drhodt} \\
&&\dot{X}_i = \sum_{j,k,l}\, N_i \left(  \Gamma_{kl \rt ij}\,
\frac{X_l^{N_l}\, X_k^{N_k}}{N_l!\, N_k !}  \; - \; \Gamma_{ij \rt
kl}\, \frac{X_i^{N_i}\, X_j^{N_j}}{N_i !\, N_j  !} \right) \equiv
\Gamma_i \vv
\label{e:dXdt} \\
&& n_B~ \sum_j Z_j\, X_j =n_{e^-}-n_{e^+}\equiv
L \left(\frac{m_e}{T}, \pe\right)  \equiv
T^3~ \lh \left(\frac{m_e}{T}, \pe\right) \vv \label{e:charneut}
\eea
where $\rho$ and $\p$ denote the total energy density and pressure,
respectively,
\bea
\rho &=& \rho_\gamma + \rho_e + \rho_\nu + \rho_B \vv \\
\p &=& \p_\gamma + \p_e + \p_\nu + \p_B\vv
\eea
while $i,j,k,l$ denote nuclear species, $N_i$ the number of
nuclides of type $i$ entering a given reaction (and analogously
$N_j$, $N_k$, $N_l$), and the $\Gamma$'s denote symbolically the
reaction rates. For example, in the case of decay of the species
$i$, $N_i=1$, $N_j=0$ and $\sum\Gamma_{i\to kl}$ is the inverse
lifetime of the nucleus $i$; for binary collisions,
$N_i=N_j=N_k=N_l=1$ and $\Gamma_{ij\to kl}=\langle \sigma_{ij\to
kl}\,v\rangle$, i.e. it represents the thermal average of the
cross section for the reaction $i+j\to k+l$ times the relative
velocity of $i$ and $j$. In Eq. (8), $Z_i$ is the charge number of
the $i-$th nuclide, and the function $\lh(\xi,\omega)$ is defined
as
\be
\lh(\xi,\omega) \equiv \frac{1}{\pi^2} \int_\xi^\infty
d\zeta~\zeta\, \sqrt{\zeta^2-\xi^2}~ \left(
\frac{1}{e^{\zeta-\omega}+1} - \frac{1}{e^{\zeta+\omega}+1}
\right) \pp \label{lfunc}
\ee
Equation~\eqn{e:drdt} is the definition of the Hubble parameter
$H$, $a$ denoting the scale factor of the
Friedmann-Robertson-Walker-Lema\^{i}tre metric, with $G_N$ the
gravitational constant, whereas Eq.s~\eqn{e:dnbdt}
and~\eqn{e:drhodt} state the total baryon number and entropy
conservation per comoving volume, respectively. The set of
$N_{nuc}$ Boltzmann equations (\ref{e:dXdt}) describes the density
evolution of each nuclide specie, with $\Gamma_{kl \rt ij}$ the
rate per incoming particles averaged over kinetic equilibrium
distribution functions. Finally, Eq.~\eqn{e:charneut} states the
Universe charge neutrality in terms of the electron chemical
potential, with $L \left(m_e/T, \pe \right)$ the charge density in
the lepton sector in unit of the electron charge.

The neutrino energy density and pressure are defined in terms of
their distributions in momentum space as
\be
\rho_\nu = 3 \, \p_\nu = 2 \, \int \frac{d^3 p}{(2 \pi)^3} \,
\left|\vec{p}\right|\, \left[ f_{\nu_e}+ 2\, f_{\nu_x} \right]\vv
\ee
Indeed, in the default scenario we assume a vanishing neutrino
chemical potential, so that $f_{\nu_e}=f_{\bar{\nu}_e}$ and $
f_{\nu_x}\equiv f_{\nu_\mu} = f_{\bar{\nu}_\mu} = f_{\nu_\tau} =
f_{\bar{\nu}_\tau}$. The nuclide evolution can be followed in \pth
also for finite neutrino chemical potential, see Section
\ref{ss:nonstandard} below.

As well known, neutrinos decouple from the electromagnetic plasma
at temperatures of a few MeV. Soon after, when the onset of
$e^{+}-e^{-}$ annihilations takes place, $e^{\pm}$ are still
partially coupled to neutrinos. The neutrino distributions are
thus slightly distorted, especially in their high energy tail (and
the $e-$flavor more than the other two, since the former also
interacts via charged current). To get BBN predictions accurate at
the sub-percent level it is necessary to follow in details this
residual out of equilibrium neutrino heating by solving the
kinetic equations for neutrino distributions. Remarkably, baryons
provide a negligible contribution to the dynamics of the Universe
at the BBN epoch as the baryon to photon number density is very
small, $\eta \alt 10^{-9}$, and therefore Boltzmann equations for
neutrino species can be solved together with
equations~\eqn{e:drdt} and~\eqn{e:drhodt} only, ignoring the
dynamics of nuclear species. This allows one to solve the
evolution of the neutrino species first, and then to substitute
the resulting neutrino distribution into the remaining equations.
We do not consider neutrino oscillations, whose effect has been
studied and shown to be sub-leading in \cite{Mangano:2005cc}. The
reader can find further details on the neutrino decoupling stage
in \cite{Serpico:2004gx,Mangano:2005cc,Mangano:2006ar}.

%%%%%%%%%%%%%%%%%%%%%%%%%%%%%%%%%%%%%%%%%%%%%%%%%%%%%%%%
\subsection{Numerical solution of the BBN set of equations}
\label{s:numsol}
%%%%%%%%%%%%%%%%%%%%%%%%%%%%%%%%%%%%%%%%%%%%%%%%%%%%%%%%
The BBN set of equations~\eqn{e:drdt}-\eqn{e:charneut} can be
recast in a form more convenient for a numerical solution, which
follows the evolution of the $N_{nuc}+1$ unknown quantities
$(\pe,~ X_j)$ as functions of the dimensionless variable
$z=m_e/T$. In this framework, Eq.~\eqn{e:charneut} provides $n_B$
as a function of $\pe$. In particular, the set of differential
equations implemented in \pth is the following:
\be
\frac{d\pe}{dz} = \frac1z \frac{\lh\, \kappa_1 +\left(\hrho_{e
\gamma B} + \hp_{e \gamma B}+
\frac{\mathcal{N}(z)}{3}\right)\,\kappa_2}{\lh\, \frac{\partial
\hrho_e}{ \partial \pe} -\frac{\partial \lh}{
\partial \pe} \left(\hrho_{e \gamma B} + \hp_{e \gamma B}+
\frac{\mathcal{N}(z)}{3}\right)}\,\,\, ,
\label{e:basic1a}
\ee
\bea
\frac{dX_i}{d z}=\dot{X}_i \frac{dt}{dz} = -\frac{\hgi}{3
z\,\hH}\, \frac{\kappa_1 \, \frac{\partial \lh}{\partial \pe}
+\kappa_2 \, \frac{\partial \hrho_{e \gamma B}}{\partial
\pe}}{\lh\,\frac{\partial\hrho_e}{\partial\pe}-\frac{\partial
\lh}{\partial \pe}\left(\hrho_{e \gamma B} + \hp_{e \gamma B} +
\frac{\mathcal{N}(z)}{3}\right) } \,\,\, ,
\label{e:basic2a}
\eea
where
\bea
\kappa_1 = 4\, \left(\hrho_{e}+ \hrho_{\gamma}\right) + \frac32~
\hp_B - z\, \frac{\partial \hrho_e}{\partial z} -z\,
\frac{\partial \hat{\rho}_\gamma}{\partial z}
+\frac{1}{\lh}\Bigg(3 \, \lh -z \frac{\partial \lh}{ \partial
z}\Bigg)\hrho_B-\frac{z^2\, \lh}{\sum_j Z_j\, X_j}\sum_i \left(
\hdmi + \frac{3}{2\, z} \right)\hgi\,\,\, ,
\eea
\bea
\kappa_2 = z\,\frac{\partial \lh}{
\partial z}-3\,\lh - z\,\lh\,\frac{ \ds \sum_i~ Z_i\, \hgi}{\ds
\sum_j~ Z_j\, X_j} \,\,\, . \eea See Appendix \ref{ap:deriv} for
notations and the explicit derivation of this set of equations.
Equations~\eqn{e:basic1a} and~\eqn{e:basic2a} are solved by
imposing the following initial conditions at $z_{in}= m_e/(10\,
{\rm MeV})$:
\bea
\pe (z_{in}) &=& \pe^0 \vv \\
X_1 (z_{in}) & \equiv & X_n (z_{in}) = \left(\exp\{\hat{q}\,
z_{in}\}+1\right)^{-1} \vv \\
X_2 (z_{in}) & \equiv & X_p(z_{in}) = \left(\exp\{-\hat{q}\,
z_{in}\}+1\right)^{-1} \vv
\\ X_i (z_{in}) &=& \frac{g_i}{2}~ \left( \zeta(3)
\sqrt{\frac{8}{\pi}} \right)^{A_i-1} ~ A_i^\frac{3}{2}\, \left(
\frac{m_e}{M_N z_{in}} \right)^{\frac{3}{2} (A_i-1)}
\eta_i^{A_i-1}\, X_p^{Z_i}(z_{in})\, \nn\\
&{\times}& X_n^{A_i-Z_i}(z_{in}) \, \exp \left\{ \hat{B}_i \,
z_{in} \right\} \quad\quad\quad\quad\quad i= \,^2{\rm H}, \,^3{\rm
H}, ... \pp
\eea
In the previous equations $\hat{q}=(M_n-M_p)/m_e$, and the
quantities $A_i$ and $\hat{B}_i$ denote the atomic number and the
binding energy of the $i-$th nuclide normalized to electron mass,
respectively. Also note that Eq. (20) is only applied if the
resulting abundance is greater than the numerical zero assumed
(variable YMIN, whose default setting is $10^{-30}$). Finally,
$\eta_i$ is the initial value of the baryon to photon number
density ratio at $T=10\,{\rm MeV}$ (for a discussion of how it is
related to the final value after $e^+-e^-$ annihilation stage see
e.g. Section 4.2.2 in \cite{Serpico:2004gx}), and $\pe^0$ the
solution of the implicit equation
\be
\lh (z_{in},\, \pe^0) =
\frac{2\, \zeta(3)}{\pi^2}~ \eta_i~ \sum_i Z_i\, X_i (z_{in}) \pp
\label{e:L}
\ee

\subsection{The Nuclear Chain}
\label{s:nucchain}

In Tables \ref{t:nucchan1}, \ref{t:nucchan2}, and \ref{t:nucchan3}
are reported the nuclear processes considered in \pth. The
enumeration shown in the first column of the tables correspond to
the order in which they appear in the program. See
\cite{Serpico:2004gx} for the relevant formalism concerning the
thermally averaged nuclear rates and an analysis of the main
experimental reaction rates. Reactions included in Table
\ref{t:nucchan1} are used when running \pth in its simpler version
(small network), while those of Tables \ref{t:nucchan2} and
\ref{t:nucchan3} are added in the intermediate and complete
network running options, which also follows the evolution of the
nuclides heavier than $^7$Be and $^{12}$N, respectively. Using the
small network gives values of the lighter nuclides like $^2$H,
$^3$He, $^4$He and $^7$Li which differ from the results obtained
with the complete network for less than 0.02\%, for default values
of the input cosmological parameters. With respect to the database
used in \cite{Serpico:2004gx}, there are a few minor upgrades
implemented here, namely the three reactions (98, 99, 100) have
been inserted following the analysis of the extended network
reported in \cite{Iocco:2007km}. Also, we have added the recent
data reported in \cite{Leonard:2006} to the regressions for the
rates (28, 29), with results in good agreement with those adopted
in \cite{Serpico:2004gx}.

\begin{table*}[t]
\begin{center}
\small
\begin{tabular}{l c c c l c c}
\hline \boldmath{No.} & ~~~~~~~~~~\boldmath{Reaction}~~~~~~~~~~ & ~~~~\boldmath{Type}~~~~ & ~~~~~ & \boldmath{No.}~~~
& ~~~~~~~~~~~~~~\boldmath{Reaction}~~~~~~~~~~~~~~ & ~~~~\boldmath{Type}~~~~\\
\hline
1 & n $\lrt$ p & \emph{weak}&& 22 & $^6$Li + p $\lrt$ $ \gamma$ + $^7$Be & (p,$\gamma$)\\
2 & $^3$H $\rightarrow$ $\bar{\nu}_e$ + $e^{-}$ + $^3$He & \emph{weak} && 23 & $^6$Li + p $\lrt$ $^3$He + $^4$He & $^3$He Pickup\\
3 & $^8$Li $\rightarrow$ $\bar{\nu}_e$ + $e^{-}$ + $2\,\, ^4$He & \emph{weak} && 24 & $^7$Li + p $\lrt$ $^4$He + $^4$He & $^4$He Pickup\\
4 & $^{12}$B $\rightarrow$ $\bar{\nu}_e$ + $e^{-}$ + $^{12}$C &
\emph{weak} && 24 bis &
$^7$Li + p $\lrt$ $ \gamma$ + $^4$He + $^4$He & (p,$\gamma$)\\
5 & $^{14}$C $\rightarrow$ $\bar{\nu}_e$ + $e^{-}$ + $^{14}$N &
\emph{weak} && 25 & $^4$He + $^2$H $\lrt$ $ \gamma$ + $^6$Li
& (d,$\gamma$)\\
6 & $^8$B $\rightarrow$ $\nu_e$ + $e^{+}$ + $2\,\,^4$He & \emph{weak} && 26 & $^4$He + $^3$H $\lrt$ $ \gamma$ + $^7$Li& (t,$\gamma$)\\
7 & $^{11}$C $\rightarrow$ $\nu_e$ + $e^{+}$ + $^{11}$B & \emph{weak} &&
27 & $^4$He + $^3$He $\lrt$ $ \gamma$ + $^7$Be
& ($^3$He,$\gamma$)\\
8 & $^{12}$N $\rightarrow$ $\nu_e$ + $e^{+}$ + $^{12}$C & \emph{weak} && 28 & $^2$H + $^2$H $\lrt$ n + $^3$He & $^2$H Strip.\\
9 & $^{13}$N $\rightarrow$ $\nu_e$ + $e^{+}$ + $^{13}$C & \emph{weak} && 29 & $^2$H + $^2$H $\lrt$ p + $^3$H & $^2$H Strip.\\
10 & $^{14}$O $\rightarrow$ $\nu_e$ + $e^{+}$ + $^{14}$N & \emph{weak} && 30 & $^3$H + $^2$H $\lrt$ n + $^4$He & $^2$H Strip.\\
11 & $^{15}$O $\rightarrow$ $\nu_e$ + $e^{+}$ + $^{15}$N & \emph{weak} && 31 & $^3$He + $^2$H $\lrt$ p + $^4$He & $^2$H Strip.\\
12 & p + n $\lrt$ $\gamma$ + $^2$H & (n,$\gamma$) && 32 & $^3$He + $^3$He $\lrt$ p + p + $^4$He& ($^3$He,$2p$)\\
13 & $^2$H + n $\lrt$ $ \gamma$ +$^3$H & (n,$\gamma$) && 33 & $^7$Li + $^2$H $\lrt$ n + $^4$He + $^4$He& (d,n $\alpha$)\\
14 & $^3$He + n $\lrt$ $ \gamma$ + $^4$He & (n,$\gamma$) && 34 & $^7$Be + $^2$H $\lrt$ p + $^4$He + $^4$He& (d,p $\alpha$)\\
15 & $^6$Li + n $\lrt $ $\gamma$ + $^7$Li & (n,$\gamma$) && 35 & $^3$He + $^3$H $\lrt$ $ \gamma$ + $^6$Li & (t,$\gamma$)\\
16 & $^3$He + n $\lrt$  p + $^3$H & charge ex. && 36 & $^6$Li + $^2$H $\lrt$ n + $^7$Be & $^2$H Strip.\\
17 & $^7$Be + n $\lrt$  p + $^7$Li & charge ex. && 37 & $^6$Li + $^2$H $\lrt$ p + $^7$Li & $^2$H Strip.\\
18 & $^6$Li + n $\lrt$ $^3$H + $^4$He & $^3$H Pickup && 38 & $^3$He + $^3$H $\lrt$ $^2$H + $^4$He & ($^3$H,d)\\
19 & $^7$Be + n $\lrt$ $^4$He + $^4$He & $^4$He Pickup && 39 & $^3$H + $^3$H $\lrt$ n + n + $^4$He& (t,n\,n)\\
20 & $^2$H + p $\lrt$ $ \gamma$ + $^3$He & (p,$\gamma$) && 40 & $^3$He + $^3$H $\lrt$ p + n + $^4$He& (t,n p)\\
21 & $^3$H + p $\lrt$ $ \gamma$ + $^4$He & (p,$\gamma$) & \\
\hline
\end{tabular}
\end{center}
\caption{The reactions used in the small network.}
\label{t:nucchan1}
\end{table*}

\begin{table*}[t]
\begin{center}
\small
\begin{tabular}{l c c c l c c}
\hline \boldmath{No.} & ~~~~~~~~~~\boldmath{Reaction}~~~~~~~~~~ & ~~~~\boldmath{Type}~~~~ & ~~~~~ & \boldmath{No.}~~~
& ~~~~~~~~~~~~~~\boldmath{Reaction}~~~~~~~~~~~~~~ & ~~~~\boldmath{Type}~~~~\\
\hline
41 & $^7$Li + $^3$H $\lrt$ n + $^9$Be & $^3$H Strip. && 58 & $^6$Li + $^4$He $ \lrt$ $ \gamma$ + $^{10}$B & ($\alpha$,$\gamma$) \\
42 & $^7$Be + $^3$H $\lrt$ p + $^9$Be & $^3$H Strip. && 59 & $^7$Li + $^4$He $\lrt$ $ \gamma$ + $^{11}$B & ($\alpha$,$\gamma$)\\
43 & $^7$Li + $^3$He $\lrt$ p + $^9$Be & $^3$He Strip.  && 60 & $^7$Be + $^4$He $\lrt$ $ \gamma$ + $^{11}$C & ($\alpha$,$\gamma$)\\
44 & $^7$Li + n $\lrt $ $\gamma$ + $^8$Li & (n,$\gamma$) && 61 & $^8$B + $^4$He $\lrt$ p + $^{11}$C & ($\alpha$,p)\\
45 & $^{10}$B + n $\lrt$ $ \gamma$ + $^{11}$B & (n,$\gamma$) && 62 & $^8$Li + $^4$He $\lrt$ n + $^{11}$B & ($\alpha$,n)\\
46 & $^{11}$B + n $\lrt$ $ \gamma$ + $^{12}$B & (n,$\gamma$) && 63 & $^9$Be + $^4$He $\lrt$ n + $^{12}$C & ($\alpha$,n)\\
47 & $^{11}$C + n $\lrt$ p + $^{11}$B & (n,p) && 64 & $^9$Be + $^2$H $\lrt$ n + $^{10}$B & ($^2$H,n)\\
48 & $^{10}$B + n $\lrt$ $^4$He + $^7$Li& (n,$\alpha$) && 65 & $^{10}$B + $^2$H $\lrt$ p + $^{11}$B & ($^2$H,p)\\
49 & $^7$Be + p $\lrt $ $\gamma$ + $^8$B & (p,$\gamma$) && 66 & $^{11}$B + $^2$H $\lrt$ n + $^{12}$C & ($^2$H,n)\\
50 & $^9$Be + p $\lrt$ $ \gamma$ + $^{10}$B & (p,$\gamma$) && 67 & $^4$He
+ $^4$He + n $\lrt$ $ \gamma$ + $^9$Be
& ($\alpha$\,n,$\gamma$)\\
51 & $^{10}$B + p $\lrt$ $ \gamma$ + $^{11}$C & (p,$\gamma$) && 68 &
$^4$He + $^4$He + $^4$He $\lrt$ $ \gamma$ + $^{12}$C
& ($\alpha$\,$\alpha$,$\gamma$)\\
52 & $^{11}$B + p $\lrt$ $ \gamma$ + $^{12}$C & (p,$\gamma$) && 69 & $^8$Li + p $\lrt$ n + $^4$He + $^4$He & (p,n\,$\alpha$)\\
53 & $^{11}$C + p $\lrt$ $ \gamma$ + $^{12}$N & (p,$\gamma$) && 70 & $^8$B + n $\lrt$ p + $^4$He + $^4$He & (n,p\,$\alpha$)\\
54 & $^{12}$B + p $\lrt$ n + $^{12}$C & (p,n) && 71 & $^9$Be + p $\lrt$ $^2$H + $^4$He + $^4$He & (p,d\,$\alpha$)\\
55 & $^9$Be + p $\lrt$ $^4$He + $^6$Li& (p,$\alpha$) && 72 & $^{11}$B + p $\lrt$ $^4$He + $^4$He + $^4$He & (p,$\alpha$\,$\alpha$)\\
56 & $^{10}$B + p $\lrt$ $^4$He + $^7$Be& (p,$\alpha$) && 73 & $^{11}$C + n $\lrt$ $^4$He + $^4$He + $^4$He & (n,$\alpha$\,$\alpha$)\\
57 & $^{12}$B + p $\lrt$ $^4$He + $^9$Be& (p,$\alpha$) &&& \\
\hline
\end{tabular}
\end{center}
\caption{The reactions used in the intermediate network in
addition to those of Table \ref{t:nucchan1}.} \label{t:nucchan2}
\end{table*}

\begin{table*}[b]
\begin{center}
\small
\begin{tabular}{l c c c l c c}
\hline \boldmath{No.} & ~~~~~~~~~~\boldmath{Reaction}~~~~~~~~~~ & ~~~~\boldmath{Type}~~~~ & ~~~~~ & \boldmath{No.}~~~
& ~~~~~~~~~~~~~~\boldmath{Reaction}~~~~~~~~~~~~~~ & ~~~~\boldmath{Type}~~~~\\
\hline
74 & $^{12}$C + n $\lrt$ $ \gamma$ + $^{13}$C & (n,$\gamma$) && 88 &
$^{12}$C + $^4$He $\lrt$ $ \gamma$ + $^{16}$O
& ($\alpha$,$\gamma$)\\
75 & $^{13}$C + n $\lrt$ $ \gamma$ + $^{14}$C & (n,$\gamma$) && 89 & $^{10}$B + $^4$He $\lrt$ p + $^{13}$C & ($\alpha$,p)\\
76 & $^{14}$N + n $\lrt$ $ \gamma$ + $^{15}$N & (n,$\gamma$) && 90 & $^{11}$B + $^4$He $\lrt$ p + $^{14}$C & ($\alpha$,p)\\
77 & $^{13}$N + n $\lrt$ p + $^{13}$C & (n,p) && 91 & $^{11}$C + $^4$He $\lrt$ p + $^{14}$N & ($\alpha$,p)\\
78 & $^{14}$N + n $\lrt$ p + $^{14}$C & (n,p) && 92 & $^{12}$N + $^4$He $\lrt$ p + $^{15}$O & ($\alpha$,p)\\
79 & $^{15}$O + n $\lrt$ p + $^{15}$N & (n,p) && 93 & $^{13}$N + $^4$He $\lrt$ p + $^{16}$O & ($\alpha$,p)\\
80 & $^{15}$O + n $\lrt$ $^4$He + $^{12}$C & (n,$\alpha$) && 94 & $^{10}$B + $^4$He $\lrt$ n + $^{13}$N & ($\alpha$,n)\\
81 & $^{12}$C + p $\lrt$ $ \gamma$ + $^{13}$N & (p,$\gamma$) && 95 & $^{11}$B + $^4$He $\lrt$ n + $^{14}$N & ($\alpha$,n)\\
82 & $^{13}$C + p $\lrt$ $ \gamma$ + $^{14}$N & (p,$\gamma$) && 96 & $^{12}$B + $^4$He $\lrt$ n + $^{15}$N & ($\alpha$,n)\\
83 & $^{14}$C + p $\lrt$ $ \gamma$ + $^{15}$N & (p,$\gamma$) && 97 & $^{13}$C + $^4$He $\lrt$ n + $^{16}$O & ($\alpha$,n)\\
84 & $^{13}$N + p $\lrt$ $ \gamma$ + $^{14}$O & (p,$\gamma$) && 98 & $^{11}$B + $^2$H $\lrt$ p + $^{12}$B & $^2$H Strip.\\
85 & $^{14}$N + p $\lrt$ $ \gamma$ + $^{15}$O & (p,$\gamma$) && 99 & $^{12}$C + $^2$H $\lrt$ p + $^{13}$C & $^2$H Strip.\\
86 & $^{15}$N + p $\lrt$ $ \gamma$ + $^{16}$O & (p,$\gamma$) && 100 & $^{13}$C + $^2$H $\lrt$ p + $^{14}$C& $^2$H Strip.\\
87 & $^{15}$N + p $\lrt$ $^4$He + $^{12}$C & (p,$\alpha$) & && \\
\hline
\end{tabular}
\end{center}
\caption{The reactions used in the complete network in addition to
those of Tables \ref{t:nucchan1} and \ref{t:nucchan2}.}
\label{t:nucchan3}
\end{table*}

%%%%%%%%%%%%%%%%%%%%%%%%%%%%%%%%%%%%
\section{Non-standard physics}\label{ss:nonstandard}
%%%%%%%%%%%%%%%%%%%%%%%%%%%%%%%%%%%%
In the standard scenario the only free parameter entering the BBN
dynamics is the value of the baryon to photon number density
$\eta$, or equivalently the baryon energy density parameter
$\Omega_B h^2$, see e.g. \cite{Serpico:2004gx} for the relation
between these parameters. If one goes beyond the standard
framework, the BBN predictions may be altered by non-standard
physics entering e.g. the
neutrino~\cite{Serpico:2005bc,Mangano:2005cc,Mangano:2006ar,Chu:2006ua}
or gravity sector~\cite{Coc:2006rt,DeFelice:2005bx}, or more
generically by the presence in the plasma of other degrees of
freedom besides the Standard Model
ones~\cite{Barger:2003rt,Cuoco:2003cu,Jedamzik:2004er,Serpico:2004nm,Cyburt:2004yc,Hansen:2001hi,Mangano:2006ur}.
For an earlier review, see~\cite{Sarkar:1995dd}. Typically,
constraints on non-minimal and/or exotic scenarios require
model-dependent modifications of the equations ruling BBN. In
{\pth} we implement a few of them described below which are
general enough to be commonly used/referred to in the specialized
literature.

%%%%%%%%%%%%%%%%%%%%%%%%%%%%%%%%%%%%
\subsection{Energy density of the vacuum, $\rho_\Lambda$}

As in the original Kawano code \cite{KawCode92}, we allow for a
non-zero cosmological constant term at the BBN epoch. We
parameterize it by means of $\rho_\Lambda$ entering the equations
only via
\be
3H\to 3\,H = \sqrt{24\,\pi\, G_N \left[\left( \frac{m_e}{z}
\right)^4 \hrho+\rhol \right]}\:.
\ee
The allowed range for this parameter in units of MeV$^4$ is
$0\leq(\rhol$/MeV$^4)\leq 1$.

%%%%%%%%%%%%%%%%%%%%%%%%%%%%%%%%%%%%
\subsection{Extra degrees of freedom, $\dneff$}

We parameterize the radiation density in non-electromagnetically
interacting particles at the BBN epoch by an additional radiation
energy density $\rho_X$ entering $H$. This is related to the
``number of extra effective neutrino species" customarily used in
the literature $\dneff$ by the equation
\be
\rho_X=\frac{7}{8}\frac{\pi^2}{30}\dneff T_X^4\,,
\ee
where, from the entropy conservation, $T_X=T=m_e/z$ at
temperatures higher than the effective neutrino decoupling
temperature, chosen as $T_d=2.3$ MeV, or else
\be
T_X=T\left[\frac{\hrho_{e,\gamma,B}(T)+
\hp_{e,\gamma,B}(T)}{\hrho_{e,\gamma,B}(T_d) +
\hp_{e,\gamma,B}(T_d)}\right]^{1/3}\,,\:\:T<T_d\:.
\ee
The user may input a value of $\dneff$ in the range
$-3.0\leq\dneff\leq 15.0$.

%%%%%%%%%%%%%%%%%%%%%%%%%%%%%%%%%%%%
\subsection{Chemical potential of the neutrinos, $\xi$}

The usual argument in favor of a cosmic lepton asymmetry is that
sphaleron effects before electroweak symmetry breaking equilibrate
the lepton and baryon asymmetries to within a factor of order
unity, thus producing the observed baryon density. In principle,
however, the electron-neutrino degeneracy parameter
$\xi=\mu_{\nu_e}/T_{\nu_e}$ as well as the degeneracy parameters
of the other neutrino flavors are not determined within the
Standard Model, and should be constrained observationally.
Recently, it has been realized that the measured neutrino
oscillation parameters imply that neutrinos reach approximate
chemical equilibrium before the BBN epoch. Thus, all neutrino
chemical potentials can be taken to be equal, i.e. they are all
characterized by the same degeneracy parameter $\xi$ that applies
to $\nu_e$ \cite{Dolgov:2002ab,Wong:2002fa,Abazajian:2002qx}. In
light of these results it is meaningful to assume a single and
shared value $\xi$ as the only free input parameter. Also, to
achieve an approximate agreement between the observed and
predicted light element abundances a possible lepton asymmetry
must be small, $|\xi|\ll 1$. For such small $\xi$ values the most
important impact on BBN is a shift of the beta equilibrium between
protons and neutrons. A subleading effect is a modification of the
radiation density,
\be
\Delta N_{\rm eff}(\xi) = 3 \bigg[ \frac{30}{7} \bigg(
\frac{\xi}{\pi} \bigg)^2 + \frac{15}{7} \bigg( \frac{\xi}{\pi}
\bigg)^4 \bigg]\,. \label{dnxi}
\ee
Moreover, the neutrino decoupling temperature is higher than in
the standard case~\cite{Freese:1982ci,Kang:1991xa}, so that in
principle one could get a non-standard $T_\nu(T)$ evolution, but
such effects are completely negligible for our case.  A non-zero
$\xi$ slightly modifies the partial neutrino reheating following
the $e^{+}e^{-}$ annihilation~\cite{Esposito:2000hi}, again a
completely negligible effect for the range of $\xi$ of our
interest. In the code, we allow the user to select among 21
possible values of $\xi$, between -1.0 and +1.0, spaced by 0.1.
The changes in the weak reactions are then automatically
implemented, as in~\cite{Esposito:2000hh}. The associated change
in $\dneff$ of Eq.~\eqn{dnxi} is also accounted for. Note that to
derive results on a finer grid a perturbative approach as the one
in \cite{Serpico:2004nm} would be required. This possibility is
left for an implementation in a future upgraded version of the
code.

%%%%%%%%%%%%%%%%%%%%%%%%%%%%%%%%%%%%
\section{The structure of \pth}
\label{s:code}
%%%%%%%%%%%%%%%%%%%%%%%%%%%%%%%%%%%%

The code is divided in two files, \texttt{main.f} and
\texttt{parthenope.f}, the former one containing the main program
and the latter the remaining subroutines. While all the physics is
implemented in \texttt{parthenope.f}, the file \texttt{main.f} is
an interface which can be possibly adapted to the user needs. The
user can choose between two running modes: an interactive one,
with parameter selections given on the screen, and a card mode
requiring an input card, an example of which is provided as the
file \texttt{input} (see also Table \ref{t:inputcard}). The
program links to the NAG libraries \cite{nag} for some algebraic
operations and the evaluation of special functions.

\begin{figure}[t]
\begin{center}
\begin{tabular}{c}
\epsfig{file=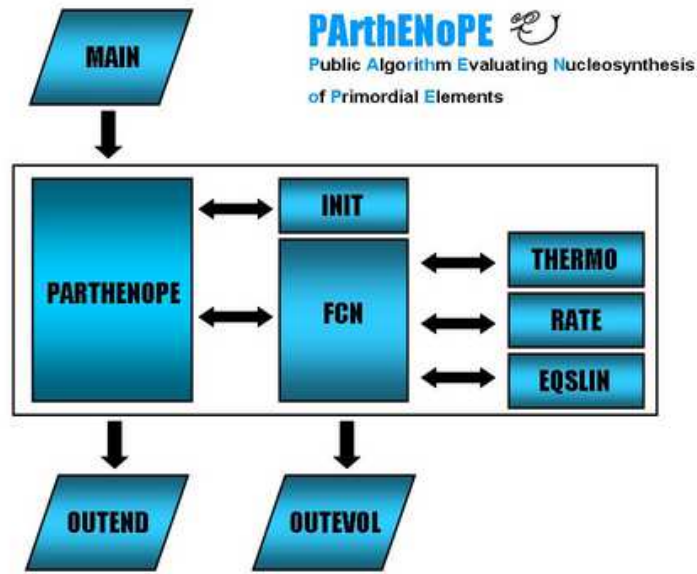,width=0.6\columnwidth}
\end{tabular}
\end{center}
\caption{The logical structure of \pth.}
\label{f:logo}
\end{figure}

The logical structure of the code is depicted in Figure
\ref{f:logo}. In the following we detail the aim of each block.

\subsection{\texttt{MAIN}}

MAIN contains the interface which allows the user to choose the
physical and network input parameters and to customize the output.

Physical parameters presently are: baryon density, number of
additional neutrino species, neutron lifetime, neutrino chemical
potential, energy density of the vacuum at the BBN epoch.
\begin{table*}[b]
\begin{center}
\begin{tabular}{lccc}
\hline
KEYWORD & DESCRIPTION & DEFAULT & RANGE/OPTIONS \\
\hline
OMEGABH   & Baryon density  $\Omega_B h^2$                          & 0.0223           & 0.01 $\div$ 0.03 \\
DNNU      & Number of additional neutrino species                   & 0.               & -3. $\div$ 15.   \\
TAU       & Neutron lifetime                                        & 885.7 s          & 880. $\div$ 890. \\
IXIE      & Integer fixing the electron neutrino chemical potential & 11               & 1 $\div$ 21      \\
RHOLMBD   & Energy density of the vacuum, $\rhol$, in MeV$^4$       & 0.               & 0. $\div$ 1.     \\
NETWORK   & Number of nuclides in the network                       & 9                & 9,18,26          \\
FOLLOW    & Option for following the evolution on the screen        & F                & T,F              \\
OVERWRITE & Option for overwriting the output files                 & F                & T,F              \\
FILES     & Name of the output files                                & parthenope.out   & 20 bit string    \\
          &                                                         & nuclides.out     & 20 bit string    \\
OUTPUT    & Evolution of nuclides                                   & first 9 nuclides & see text         \\
RATES     & Details on changed rates                                & No change        & see text         \\
EXIT      & Closing keyword in the input card                       &                  &                  \\
\hline
\end{tabular}
\end{center}
\caption{The list of the possible keywords in the input card,
their default values and corresponding ranges/options.}
\label{t:keywords}
\end{table*}

Network parameters include: the choice among a small (9 nuclides
and 40 reactions), an intermediate (18 nuclides and 73 reactions),
and a complete network (26 nuclides and 100 reactions). Moreover,
the user can change the rates of each reaction included in the
chosen network, selecting a `LOW' or a `HIGH' value, based on the
experimental or theoretical uncertainties, or a customized
multiplicative `FACTOR'.

Finally, the output options include: the choice of the nuclides
whose evolution has to be followed versus $z$, the name of the
output files (a first one with the final results and a second with
the evolution of the selected nuclides), and the possibility to
follow the status of the evolution on the screen.

All this information can be provided either interactively,
following the on-screen instructions, or by an input card, with
the format of the example card included in the distribution. In
particular, each line in this card must start with an allowed key
and the last line key has to be `EXIT'. In order to be recognized,
each key must start at the first bit of the line. The allowed
keywords are listed in Table \ref{t:keywords}, together with the
default values adopted by the code whenever the corresponding key
is not explicitly set.

An example of input card is shown in Table \ref{t:inputcard}.
While the keys OMEGABH, DNNU, TAU, IXIE, RHOLMBD, NETWORK, FOLLOW,
and OVERWRITE have only one argument, the keyword FILES has two
arguments, that is the two names of the output files, each one at
most 20 bits long. Some more details deserve the two keywords
OUTPUT and RATES. OUTPUT can have at most $N_{nuc}+2$ arguments
($N_{nuc}$ being the number of nuclides of the chosen network),
which are: 1) a bit equal to `T' or `F', if the user wants or not
to store in the output the evolution of a given set of nuclide
abundances, 2) the total number of such nuclides, and 3) the
identity of these nuclides (given as the corresponding number in
Table \ref{t:nuclnumb}). In the example of Table \ref{t:inputcard}
with the sequence T 3 2 3 4 the user has chosen to store in the
output the three nuclides p, $^2$H and $^3$H. The keyword RATES
allows to change the default values of the nuclear rates used in
the chosen network. The input card can have more than one line
with this keyword, as in the example of Table \ref{t:inputcard}.
Each line contains: 1) an integer k, giving the number of
reactions whose change is specified on that line; 2) the kind of
change for the k reactions with the syntax \texttt{(m i f)},
indicating that the reaction number \texttt{m} (see Tables
\ref{t:nucchan1}, \ref{t:nucchan2} and \ref{t:nucchan3}) has to be
changed according to the type of change \texttt{i}, with the
factor \texttt{f}. Obviously, \texttt{m} can assume the values
1,...,M (M being the number of reactions of the chosen network),
while \texttt{i}=1,2,3 corresponds to the `LOW', `HIGH', or
`FACTOR' type of change, respectively. Whenever a statistically
sound analysis is possible, as it is the case for most of the main
reactions, the `LOW'/`HIGH' rates represent 1 $\sigma$ lower/upper
limits to the rate, as compiled in \cite{Serpico:2004gx}. For most
of the subleading reactions, they represent estimated ranges of
variability, obtained from the literature. The main (or unique)
reference from which the rate has been taken is reported as a
comment in the code next to the related reaction line. Finally, if
\texttt{i}=3 the real number \texttt{f} is the value of the
multiplicative factor applied to the chosen reaction rate (not
considered if the options \texttt{i}=1,2 are selected). For
example, the first line of the input card of Table
\ref{t:inputcard} specifies that 3 reaction rates should be
changed in running \pth as follows
\bea
{\rm p} + {\rm n} \lrt \gamma + ^2\!{\rm H} && {\rm low \,\, rate}
\nonumber \\
^2{\rm H} + ^2\!{\rm H} \lrt {\rm n} + ^3\!{\rm He} && {\rm
rate\,\,
multiplied\,\, by\,\, the\,\, factor \,\, 0.4} \nonumber \\
^2{\rm H} + ^2\!{\rm H} \lrt {\rm p} + ^3\!{\rm H} && {\rm high
\,\, rate} \nonumber
\eea
Notice that it is possible to add comments after the parameters in
the input card and the order of the lines with different keywords
is not important.

\begin{table}[t]
\begin{center}
\begin{tabular}{l l l}
\hline
RATES      &  3  ( 12 1 0. ) (  28 3 .4) (29 2 0 ) & options for changing the nuclear rates \\
RATES      &  2  ( 3 2 0. ) (5 3 .6)            & options for changing the nuclear rates \\
TAU        &  885.7                             & experimental value of neutron lifetime \\
DNNU       &  .0                                & number of extra neutrinos \\
IXIE       & 11                                 & integer giving the value of $\nu_e$ chemical potential \\
RHOLMBD    & .0                                 & value of cosmological constant energy density at the BBN epoch \\
OVERWRITE  & T                                  & option for overwriting the output files \\
FOLLOW     & T                                  & option for following the evolution on the screen \\
OMEGABH    & .0223                              & value of $\Omega_B h^2$ \\
NETWORK    & 9                                  & number of nuclides in the network \\
FILES      & parthenope1.out nuclides1.out~~~~~ & names of the two output files \\
OUTPUT     & T  3  2 3 4                        & options for customizing the output \\
EXIT       &                                    & terminates input \\
\hline
\end{tabular}
\end{center}
\caption{An example of input card.}
\label{t:inputcard}
\end{table}

\subsection{\texttt{PARTHENOPE}}

This subroutine drives the resolution of the BBN set of equations.
It starts calling the initialization routine INIT, then the NAG
solver, finally the output printing routine OUTEND. The NAG
resolution parameters, controlling for example the resolution
method and the numerical accuracy, have been chosen to optimize
the performances of the NAG solver. Any change of these parameters
should be implemented only after a careful reading of the NAG
manual \cite{nag}.

Further relevant parameters are \texttt{zin} and \texttt{zend},
setting respectively the initial and final value of the
independent variable $z$. Their present values correspond to the
two temperatures of $T_{i} = 10\,$MeV and $T_{f} = 1/130\,$MeV.
Note that a few settings depend on these values, which then should
be varied with caution.

\subsection{\texttt{INIT}}

Besides initializing the nuclear parameters, this subroutine
calculates the initial values for all nuclide abundances and the
electron chemical potential, the latter requiring the inversion of
the implicit equation~\eqn{e:L} with a NAG routine.

\subsection{\texttt{FCN, THERMO, RATE, EQSLIN}}

The subroutine \texttt{FCN} is required by the NAG solver to
calculate the right hand side of the differential BBN equations.
In order to do this, the thermodynamical quantities which appear
in the equations are evaluated with a call to the subroutine
\texttt{THERMO}. The second step is the calculation of the
reaction rates with the subroutine \texttt{RATE}. Then the
linearization of the set of equations is performed, with the
construction of a corresponding $N_{nuc}\times N_{nuc}$ matrix
($N_{nuc}$ being the number of nuclides). In this way, the unknown
functions appear in a linear equation system, solved by Gaussian
elimination in the subroutine \texttt{EQSLIN} (this method is very
similar to the one applied in the Kawano code \cite{KawCode92}).

\subsection{\texttt{OUTEVOL, OUTEND}}

The subroutine \texttt{OUTEVOL} is called during the evolution,
for printing the intermediate values of the chosen nuclide
abundances in one of the output files. Moreover, if requested by
the user, this subroutine allows to follow the resolution
evolution, printing some physical quantities on the screen.
Finally, \texttt{OUTEND} prints the final values of the nuclide
abundances and electron chemical potential in the other output
file along with some technical information on the differential
evolution resolution. The final yield of the $i-$th nuclide is
expressed as the ratio $X_{i}/X_p$, i.e. number density normalized
to hydrogen. The only exceptions are Hydrogen expressed as $X_p$
and $^4$He, which is conventionally reported in terms of the
(approximate) mass fraction $Y_p = 4 X_{^4{\rm He}}$.

%%%%%%%%%%%%%%%%%%%%%%%%%%%%%%%%%%%%%%%%
\section{Main differences with respect to the Wagoner-Kawano
code}
\label{s:compare}
As we have previously emphasized, the public Kawano code
\cite{KawCode92} was the starting point for the development of
\pth. So, they have a similar structure, like the subdivision in
several subroutines which contain, for example, the interface with
the user, the calculation of nuclear rates, of the thermodynamical
quantities, the differential equation solver, and the production
of the output. Here we summarize the main physical and numerical
differences with respect to the original code:
\begin{itemize}
\item The interface menu and the options are now different. In
particular, the user can more easily implement changes in the
nuclear reaction network. A card-mode input is available. The
output is easily customized. \item Several numerical routines have
been replaced with more efficient algorithms, most of which using
NAG routines. \item Improved calculations for the $n-p$ reactions
are implemented via new fits, not as effective corrections added a
posteriori. They also include effects of finite nucleon mass and
non-thermal neutrino spectral distortions. The same holds for the
case with neutrino asymmetry. \item An improved and more accurate
calculation of thermodynamical variables is implemented, both in
the electromagnetic sector and the neutrino one. In particular,
for the latter case we do not simply impose entropy conservation,
but take into account entropy transfer during the $e^+-e^-$
annihilation phase. \item We implemented an updated nuclear
reaction network, obtained using new data and reduction
techniques.  Also, new reactions and new {\it types} of reactions
(different number of nuclides in the initial and final states) are
now included. The new code has been restructured so that it is
easier to implement new reactions and that only fundamental
nuclear data are needed as input. All derived quantities are
calculated in the code. This is aimed at simplifying future
updates. \item We improved the numerical resolution of the coupled
BBN equations, by using a multistep method, belonging to the class
of Backward Differentiation Formulas (instead of the traditional
Runge-Kutta solver of the Kawano code which is a single
step-method) implemented by a NAG routine. The default values of
the accuracy parameters have been chosen to guarantee a good
compromise between the accuracy goal and a reasonable running
time. In particular the relative accuracy reached on $^4$He mass
fraction is of the order of $10^{-4}$, thus keeping the numerical
error below the level of theoretical uncertainties.
\end{itemize}

\begin{figure}[t]
\begin{center}
\begin{tabular}{c}
\epsfig{file=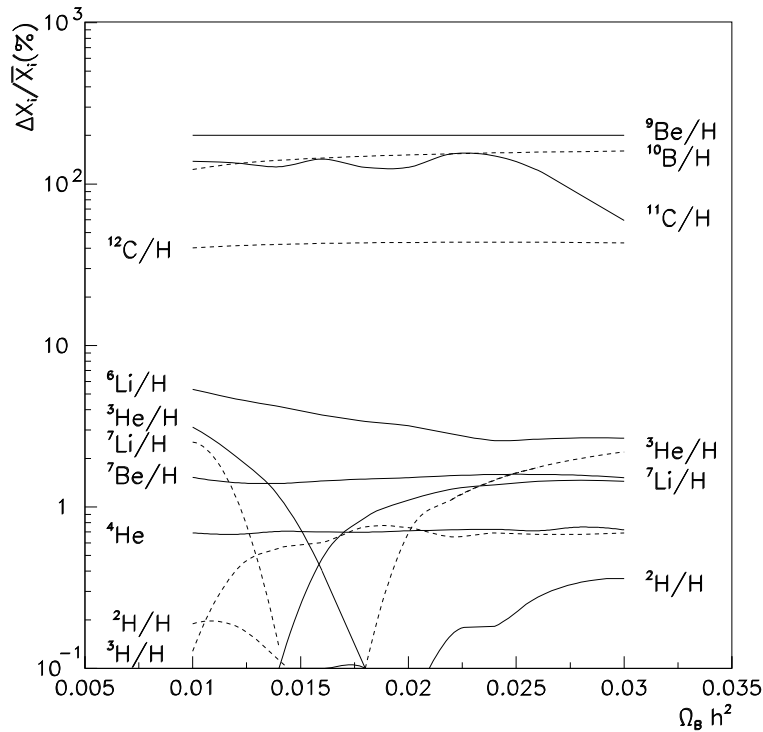,width=0.6\columnwidth}
\end{tabular}
\end{center}
\caption{The relative difference in percent of nuclei
abundances between \pth and the Kawano code, $\Delta X_i/ \bar X_i \equiv
2 (X_i^P-X_i^K)/ (X_i^P+X_i^K)$, versus baryon density $\Omega_B h^2$. The
solid (dashed) curves are for positive (negative) values.}
\label{f:comparison}
\end{figure}

In Figure \ref{f:comparison} we show the relative variation (in
percent) of nuclear abundances computed with \pth and using the
original Kawano code vs. baryon density, for the fiducial values
of all the other parameters. To give an idea of the differences
introduced by accounting for the major updates to the {\it
physics} implemented in the code, in Table \ref{t:diff} we report,
for $\Omega_B h^2=0.022$ (i.e. consistent with the value singled
out by the WMAP Collaboration \cite{Spergel:2006hy}), the relative
variation with respect to the Kawano code results of the
theoretical prediction of abundances in three cases: 1) improved
treatment of n-p weak rates, as presently in {\texttt{PArthENoPE},
but original nuclear network as in Kawano code (first column); 2)
improved treatment of n-p weak rates as presently in \pth and
updated nuclear network as in Ref. \cite{Serpico:2004gx} (second
column); 3) \pth code, with improved calculation of
thermodynamical variables, both in the electromagnetic sector and
the neutrino one, and complete improvement of the nuclear network
(third column). In summary, the \pth prediction for $^4$He differs
with respect to the original Kawano code  by 0.7 $\%$, mainly due
to the improved treatment of radiative corrections, finite
temperature and finite nucleon mass corrections in neutron--proton
weak rates. Actually, this value is larger than the theoretical
accuracy of \pth on $Y_p$, which is of the order of $0.2 \%$, see
also \cite{Serpico:2004gx}. On the other hand, our results are,
for example, in very good agreement with \cite{Fiorentini:1998fv},
where all mentioned corrections to weak rates are included as
discussed in details in \cite{Sarkar:1995dd}. Rescaling the value
of neutron lifetime to the values $\tau_n=886.7$ s adopted in
\cite{Fiorentini:1998fv} and using their fit for $Y_p$ the
agreement is at the $0.1 \%$ level. Concerning the other nuclides,
the variation of the theoretical predictions with respect to the
Kawano code reaches larger values, of the order of 3 \% for
$^6$Li, at the 1\% level for $^7$Li and $^7$Be, while it is very
small for $^2$H, as in this case the effects of improvements on
weak rates treatment and updated nuclear rate network have
different sign with respect to the one in plasma and neutrino
treatment and almost cancel out accidentally. We notice that the
introduction of new processes (reactions 98, 99, 100 of Table
\ref{t:nucchan3}), see \cite{Iocco:2007km}, and the update of
those already considered in the Kawano code result in a large
difference in the BBN theoretical prediction for metallicity,
although probably insufficient to change the chemistry of
primordial clouds. Finally, the reader may want to consider
similar comparisons performed in the literature between updated
versions of BBN codes used, e.g. in~\cite{Cyburt:2001pp},
\cite{Nollett:2000fh} and the already quoted
\cite{Fiorentini:1998fv}, and the results of the Kawano code
presented in~\cite{Smith:1992yy}.

\begin{table*}[t]
\begin{center}
\begin{tabular}{lccc}
\hline \boldmath{Nuclide} & ~~\boldmath{weak\,rates (\%)}~~ & \boldmath{nucl.\,rates (\%)} & ~~\boldmath{Parthenope (\%)}~~ \\
\hline $^2$H  &1.5   &3.2   &0.2   \\
\hline $^3$H  &3.3   &2.3   &-0.7    \\
\hline $^3$He &0.1   &-3.4    &-1.1    \\
\hline $^4$He &0.4   &0.4   &0.7   \\
\hline $^6$Li &-2.7    &12.2  &2.8   \\
\hline $^7$Li &-1.8    &2.9   &1.3   \\
\hline $^7$Be &-2.0    &2.5   &1.6   \\
\hline $^9$Be &-0.1    &200.0 &200.0 \\
\hline $^{10}$B &1.3   &-150.0  &-153.8  \\
\hline $^{11}$C &200.0 &146.0 &153.8 \\
\hline $^{12}$C &0.4   &0.5   &-43.6   \\
\hline
\end{tabular}
\end{center}
\caption{The relative variation in percent of the theoretical
predictions between \pth and the Kawano code, $\Delta X_i/ \bar X_i \equiv
2 (X_i^P-X_i^K)/ (X_i^P+X_i^K)$, is shown for three cases (see text). The
results are shown for $\Omega_B h^2=0.022$ and standard values for all
other physical parameters. }
\label{t:diff}
\end{table*}

\section{Conclusions}
\label{s:concl}

In this paper we have described the general structure and features
of \pth, a new numerical code which computes the theoretical
abundances of nuclei produced during BBN, as function of several
input cosmological parameters. This code has been recently made
public and can be obtained at the URL
http://parthenope.na.infn.it/. The code evaluates the abundances
of 26 nuclide in the standard BBN scenario, as well as in extended
models allowing for extra relativistic particles or neutrino
chemical potential. We checked for a few fiducial cases that the
results for the $^4$He abundance agree very closely with those of
Ref. \cite{Fiorentini:1998fv}, where the appropriate corrections
\cite{Sarkar:1995dd} are applied to the original Wagoner/Kawano
code \cite{Wagoner,KawCode92}.

In view of the improved data coming from astrophysical
observations, accurate tools providing theoretical predictions on
cosmological observables are required to check the overall
consistency of the picture of the evolution of the Universe, as
well as for investigating and constraining new physics beyond the
present framework of fundamental interactions.

Much effort has been put in the recent years by several groups in
order to increase the level of accuracy of theoretical prediction
on nuclide abundances, in particular by improving the estimate of
the neutron to proton weak conversion rates and the nuclear
network rates. The results of these studies, along with a to date
analysis of experimental results on relevant nuclear reactions
have been implemented in \pth, which hopefully will turn useful as
an accurate tool for BBN-related studies.

\bigskip

{\bf Acknowledgments}\\ In Naples, this work was supported in part
by the PRIN04 ``Fisica Astroparticellare e Cosmologia" and PRIN06
``Fisica Astroparticellare: neutrini ed universo primordiale" of
the Italian MiUR. P.D.S. acknowledges support by the US Department
of Energy and by NASA grant NAG5-10842.

\appendix

\section{Derivation of the \pth set of equations}
\label{ap:deriv}

We define $z\equiv m_{e}/T$, $x=m_{e}\,a$, $\bar{z}=x/z=a
T=T/T_\nu$, $\hnb=m_e^{-3}\,n_B$ and introduce the following
quantities:
\bea
&\mathcal{N}(z)=\frac{1}{\bar{z}^4}\left(x
\frac{d}{dx}\bar\rho_\nu\right)\Bigg|_{x=x(z)}\,\,\,,  &
\bar{\rho}_{\nu}=a^4\,\rho_{\nu}=\left(\frac{x}{m_e}\right)^4\,\rho_{\nu}
\,\,\,,
\label{Nz}\\
&\rho=\rho_{e \gamma B}+\rho_\nu \,\,\,,  &\p=\p_{e \gamma B}+\p_\nu \,\,\,, \\
&\hrho=T^{-4}\,\rho=\left(\frac{z}{m_e}\right)^4\,\rho \,\,\,,
&\hp=T^{-4}\,\p=\left(\frac{z}{m_e}\right)^4\,\p \,\,\,. \\
\eea
Starting from Eq.s~\eqn{e:dnbdt} and~\eqn{e:drhodt},
\bea
&&\frac{\dot{n}_B}{n_B} = -\, 3\, H \,\,\,,
\label{eq2} \\
&&\dot{\rho} = -\, 3 \, H~ (\rho + \p) \,\,\,, \label{eq1}
\eea
and separating the neutrino contribution one gets
\bea
\dot{\rho}_{e \gamma B}+\dot{\rho}_\nu =-\,3\,H\,(\rho_{e \gamma
B} + \p_{e \gamma B})-\,4\,H\,\rho_{\nu} \,\,\, , \label{drho}
\eea
where in~\eqn{drho} we have used $\rho_{\nu}=3\,\p_{\nu}$. From
Eq.~\eqn{eq1} one gets the time derivative
\be
\dot{\hrho}=\left(\frac{z}{m_e}\right)^4
\dot{\rho}+4\left(\frac{z}{m_e}\right)^3\frac{\dot{z}}{m_e}\rho
\,\,\,,
\ee
and  thus
\bea
\dot{\hrho}_{e \gamma B} = -\,3\,H\,(\hrho_{e \gamma
B} + \hp_{e \gamma B})+\,4\,\frac{\dot{z}}{z}\hrho_{e \gamma
B}-\left(\frac{z}{m_e}\right)^4(\dot{\rho_\nu}+\,4\,H\,\rho_\nu)
\label{drhoem} \,\,\, .
\eea
For the neutrino energy density one gets
\bea
\dot{\rho}_\nu=\frac{d\rho_\nu}{dx} \dot{x}=m_e \dot{a}
\frac{d\rho_\nu}{dx} = H x
\frac{d\rho_\nu}{dx}&=&\left(\frac{m_e}{x}\right)^4 H \left[x
\frac{d\bar{\rho}_\nu}{dx}-4 \bar{\rho}_\nu\right]\,\,\,, \nonumber\\
\dot{\rho}_\nu+\,4\,H\,\rho_\nu &=& \left(\frac{m_e}{x}\right)^4 H
x \frac{d\bar{\rho}_\nu}{dx} \label{drhonu}\,\,\, .
\eea
Hence substituting (\ref{drhonu}) in (\ref{drhoem}) we obtain
\be
\dot{\hrho}_{e \gamma B}=4\,\frac{\dot{z}}{z}\,\hrho_{e \gamma
B}-\,3\,H\,(\hrho_{e \gamma B} + \hp_{e \gamma
B})-H\,\mathcal{N}(z) \,\,\, .\label{dotrho1}
\ee
On the other hand, the total time derivative of $\hrho_{e \gamma
B}$ can be expressed via the partial derivatives with respect to
$z$, $\pe$ and $X_i$\,,
\be
\dot{\hrho}_{e \gamma B} = \frac{\partial \hrho_{e \gamma B}}
{\partial z} \, \dot{z} + \frac{\partial \hrho_{e \gamma B}}
{\partial \pe} \, \dot{\pe} +\sum_i \frac{\partial \hrho_{e \gamma
B}}{\partial {X_i}}\, \dot{X}_i = \left( \frac{\partial \hrho_{e
\gamma B}}{\partial z} + \frac{\partial \hrho_{e \gamma
B}}{\partial \pe} \, \frac{d\pe}{dz} +\sum_i \frac{\partial
\hrho_{e \gamma B}}{\partial {X_i}} \, \frac{dX_i}{dz} \right)
\dot{z}\,\,\,. \label{dotrho2}
\ee
Thus, equating the r.h.s of~\eqn{dotrho1} and \eqn{dotrho2} after
some rearrangement reads
\be
\left( \frac{\partial \hrho_{e \gamma B}}{\partial z} -
\frac{4}{z}\,\hrho_{e \gamma B} + \frac{\partial \hrho_{e \gamma
B}}{\partial \pe} \, \frac{d\pe}{dz}  +\sum_i \frac{\partial
\hrho_{e \gamma B}}{\partial {X_i}} \, \frac{dX_i}{dz} \right)
\dot{z} =-\,3\,z\,H\,(\hrho_{e \gamma B} + \hp_{e \gamma
B})-H\,\mathcal{N}(z)\,\,\, . \label{dotz1}
\ee
Starting from~\eqn{eq2} and proceeding in the same way leads to
\be
\left(\frac{\partial \hnb} {\partial z}+ \frac{\partial \hnb}
{\partial \pe} \frac{d\pe}{dz}+\sum_i \frac{\partial \hnb}
{\partial {X_i}} \frac{dX_i}{dz} \right) \dot{z} = - \, 3 \, z \,
H \, \hnb \,\,\, .
\label{dotz2}
\ee
Obtaining $\dot{z}$ from~\eqn{dotz2} and substituting
into~\eqn{dotz1} we obtain
\be
-\,3\,H\,\hnb\frac{ \frac{\partial \hrho_{e \gamma B}}{\partial
z}-\frac{4}{z}\hrho_{e \gamma B}+ \frac{\partial \hrho_{e \gamma
B}}{\partial \pe}\,\frac{d\pe}{dz} +\sum_i \frac{\partial \hrho_{e
\gamma B}}{\partial {X_i}} \,\frac{dX_i}{dz}}{ \frac{\partial
\hnb}{\partial z}+\frac{\partial \hnb}{\partial \pe}
\frac{d\pe}{dz}+\sum_i \frac{\partial \hnb}{\partial {X_i}}
\frac{dX_i}{dz}}=-\,3\,z\,H\,(\hrho_{e \gamma B} + \hp_{e \gamma
B})-H\,\mathcal{N}(z) \,\,\, .
\label{eq3}
\ee
By using Eq.~\eqn{e:charneut}
\be
\hnb = \frac{\lh(z,\pe)}{z^3 \sum_i Z_i X_i}\,\,\,,
\ee
one can express $\hnb$ and its derivatives as function of
$\lh(z,\pe)$ which is defined in (\ref{lfunc}). By solving
Eq.~\eqn{eq3} with respect to $d\pe/dz$ one gets
\be
\frac{d\pe}{dz} = \frac1z \frac{\lh\, \kappa_1 +\left(\hrho_{e
\gamma B} + \hp_{e \gamma B}+
\frac{\mathcal{N}(z)}{3}\right)\,\kappa_2}{\lh\, \frac{\partial
\hrho_e}{ \partial \pe} -\frac{\partial \lh}{
\partial \pe} \left(\hrho_{e \gamma B} + \hp_{e \gamma B}+
\frac{\mathcal{N}(z)}{3}\right)}\,\,\, ,\label{dphidz}
\ee
where
\bea
\kappa_1 = 4\, \left(\hrho_{e}+ \hrho_{\gamma}\right) + \frac32~
\hp_B - z\, \frac{\partial \hrho_e}{\partial z} -z\,
\frac{\partial \hat{\rho}_\gamma}{\partial z}
+\frac{1}{\lh}\Bigg(3 \, \lh -z \frac{\partial \lh}{ \partial
z}\Bigg)\hrho_B-\frac{z^2\, \lh}{\sum_j Z_j\, X_j}\sum_i \left(
\hdmi + \frac{3}{2\, z} \right)\hgi\,\,\, ,
\eea
\bea
\kappa_2 = z\,\frac{\partial \lh}{\partial z}-3\,\lh -
z\,\lh\,\frac{ \ds \sum_i~ Z_i\, \hgi}{\ds \sum_j~ Z_j\, X_j}
\,\,\, .
\eea
According to our notations, $\hdmi$ and $\hmu$ stand for the i-th
nuclide mass excess and the atomic mass unit, respectively,
normalized to $m_e$, whereas $H \equiv m_e\, \hH$ and $\Gamma_i
\equiv m_e\, \hgi$. By substituting in (\ref{dotz2}) the
expression obtained for $d\pe/dz$ in (\ref{dphidz}) we get
\bea
\dot{z}=-\,3\,H\frac{\frac{\partial
\hnb}{\partial_\pe}\left(\hrho_{e \gamma B} + \hp_{e \gamma B} +
\frac{\mathcal{N}(z)}{3}\right)-\hnb\, \frac{\partial \hrho_{e
\gamma B}}{\partial \pe}}{\frac{\partial
\hnb}{\partial_\pe}\left(\frac{\partial \hrho_{e \gamma
B}}{\partial z}-\frac4z \hrho_{e \gamma B}+\sum_i \frac{\partial
\hrho_{e \gamma B}}{\partial {X_i}}\,
\frac{dX_i}{dz}\right)-\frac{\partial \hrho_{e \gamma B}}{\partial
\pe}\left(\frac{\partial \hnb}{\partial z}+\sum_i \frac{\partial
\hnb}{\partial {X_i}}\, \frac{dX_i}{dz}\right)}\,\,\,,
\eea
namely
\bea
\frac{dt}{dz}= -\frac{\kappa_1 \, \frac{\partial \lh}{\partial
\pe} +\kappa_2 \, \frac{\partial \hrho_{e \gamma B}}{\partial
\pe}}{3\,H\,\left[\hnb\frac{\partial \hrho_{e \gamma B}}{\partial
\pe}-\frac{\partial \hnb}{\partial \pe}\left(\hrho_{e \gamma B} +
\hp_{e \gamma B} + \frac{\mathcal{N}(z)}{3}\right)\right]}\,\,\,.
\eea
The equations for the abundances~\eqn{e:dXdt} then become
\bea
\frac{dX_i}{d z}=\dot{X}_i \frac{dt}{dz} = -\frac{\hgi}{3
z\,\hH}\, \frac{\kappa_1 \, \frac{\partial \lh}{\partial \pe}
+\kappa_2 \, \frac{\partial \hrho_{e \gamma B}}{\partial
\pe}}{\lh\,\frac{\partial\hrho_e}{\partial\pe}-\frac{\partial
\lh}{\partial \pe}\left(\hrho_{e \gamma B} + \hp_{e \gamma B} +
\frac{\mathcal{N}(z)}{3}\right) }\,\,\, .
\eea
The solution of neutrino dynamics performed in
\cite{Mangano:2001iu,Mangano:2005cc,Mangano:2006ar} allows to
compute the quantity
\be
\mathcal{N}(z) = \left. \frac{1}{\bar{z}^4} \left( x \frac{d}{dx}
\bar{\rho}_\nu \right) \right|_{x=x(z)} \pp \label{e:funcf}
\ee
Notice that $\mathcal{N}(z)$ would vanish for purely thermal
neutrinos, and it is strictly related to the small entropy
transfer to neutrinos during the $e^+-e^-$ annihilation stage. In
the code, $\mathcal{N}(z)$ is calculated by using the following
fit, which is accurate to better than 0.1\% in the relevant range:
\bea
\mathcal{N}(z) &=& \left\{ \begin{array} {cc}
\exp{\left(\sum_{l=1}^{13} n_l
~z^{l}\right)} & z<4 \vv \\
 0 & z\geq 4 \vv \end{array} \right.
\eea
with
\bea
\begin{array}{ll}
n_0= -10.21703221236002 & n_1= 61.24438067531452 \\ n_2=
-340.3323864212157 &
n_3= 1057.2707914654834 \\ n_4= -2045.577491331372 & n_5= 2605.9087171012848 \\
n_6= -2266.1521815470196 & n_7= 1374.2623075963388 \\ n_8=
-586.0618273295763 &
n_9= 174.87532902234145 \\ n_{10}= -35.715878215468045 & n_{11}= 4.7538967685808755 \\
n_{12}= -0.3713438862054167 & n_{13}= 0.012908416591272199 \pp
\end{array} \nn\\
\eea
\newpage
\section{Common variables used in \pth}
\begin{center}
\begin{tabular}{|l|l|l|}
\hline
VARIABLE~~~~~~~~~~~~~~ & DESCRIPTION~~~~~~~~~~~~~~~~~~~~~~~~~~~~~~~~~~~~~~~~~~~~~~~~~~~~~~~~~~~~~~~~~~~~~~~~~ & COMMON~~~~~~~ \\
\hline
AA(NNUC)         & Nuclide atomic numbers, $A_i$ & ANUM \\
\hline
FACTOR(NREC)     & Multiplicative factor for the rate of reaction i-th & CHRATE \\
HCHRAT(NREC)     & Type of changes adopted for reaction i-th & \\
NCHRAT           & Number of reactions to be changed & \\
WCHRAT(NREC)     & Reactions to be changed & \\
\hline
ALF              & Fine structure constant, $\alpha$ & CONSTANTS \\
COEF(4)          & Unity conversion factors & \\
GN               & Newton constant, $G_N$, in MeV$^{-2}$ & \\
ME               & Electron mass, $m_e$ & \\
MU               & Atomic mass unit, $M_u$ & \\
PI               & $\pi$ & \\
\hline
IFCN             & Counter & COUNTS \\
IFCN1            & Counter & \\
ISAVE1           & Counter & \\
ISAVE2           & Counter & \\
ISTEP            & Counter & \\
\hline
DZ               & Stepsize of the independent variable in the resolution of the & DELTAZ \\
                 & nucleosynthesis equations & \\
DZ0              & Initial value for DZ & \\
\hline
DM(0:NNUC)       & Mass excesses in MeV, $\Delta M_i$ & DMASS \\
\hline
DMH(NNUC)        & Adimensional mass excesses, $\Delta M_i/m_e$ & DMASSH \\
\hline
PHI              & Adimensional electron chemical potential, $\phi_e$ & ECHPOT \\
\hline
AG(NNUC,4)       & Nuclear partition function coefficients & GPART \\
\hline
YY0(NNUC+1)      & Initial values of electron chemical potential, $\phi_e$, and nuclide & INABUN \\
                 & abundances, $X_i$ & \\
\hline
CMODE            & Flag for the choice of the running mode & INPCARD \\
FOLLOW           & Option for following the evolution on the screen (card mode) & \\
OVERW            & Option for overwriting the output files (card mode) & \\
\hline
INC              & Maximum value of the flag for the convergence of the matrix inversion & INVFLAGS \\
MBAD             & Error flag for the matrix inversion & \\
\hline
LH0              & Initial value for the adimensional electron/positron asymmetry, $\lh$ & INVPHI \\
Z0               & Initial value for the evolution variable $z=m_e/T$ (=ZIN) & \\
\hline
AMAT(NNUC,NNUC)  & Matrix involved in the linearization of the relation between $X_i(z+dz)$ & LINCOEF \\
                 & and $X_i(z)$ & \\
BVEC(NNUC)       & Vector involved in the linearization of the relation between $X_i(z-dz)$ & \\
                 & and $X_i(z)$ (contains $X_i$ in reverse order) & \\
YX(NNUC)         & $X_i(z)$ in reverse order & \\
\hline
YMIN             & Numerical zero of nuclide abundances & MINABUN \\
\hline
DNNU             & Number of extra effective neutrinos, $\Delta N_{eff}$ & MODPAR \\
DNNUXI           & Contribution to the number of extra effective neutrinos from a non & \\
                 & zero neutrino chemical potential, $\Delta N_{eff}$ of Eq.~\eqn{dnxi} & \\
ETAF             & Final value of the baryon to photon density ratio, $\eta_f$ & \\
IXIE             & A positive integer fixing the electron neutrino chemical potential & \\
RHOLMBD          & Energy density corresponding to a cosmological constant, $\rho_\Lambda$ & \\
TAU              & Value of neutron lifetime in seconds, $\tau_n$ & \\
XIE              & Electron neutrino chemical potential, $\xi$ (=XIE0) & \\
\hline
XIE0(NXIE)       & Electron neutrino chemical potential, $\xi$ & NCHPOT \\
\hline
INUC             & Number of nuclides in the selected network & NETWRK \\
IREC             & Number of reactions among nuclides in the selected network & \\
IXT(30)          & Code of the nuclides whose evolution has to be followed & \\
                 & (ixt(30)=control integer) & \\
NVXT             & Number of nuclides whose evolution has to be followed & \\
\hline
\end{tabular}
\end{center}

\begin{center}
\begin{tabular}{|l|l|l|}
\hline
VARIABLE~~~~~~~~~~~~~~ & DESCRIPTION~~~~~~~~~~~~~~~~~~~~~~~~~~~~~~~~~~~~~~~~~~~~~~~~~~~~~~~~~~~~~~~~~~~~~~~~~ & COMMON~~~~~~~ \\
\hline
BYY(NNUC+1)      & Text strings for the output & NSYMB \\
\hline
MN               & Neutron mass, $M_n$ & NUCMASS \\
MP               & Proton mass, $M_p$ & \\
\hline
NAMEFILE1        & Name of the output file for the final values of the nuclide abundances & OUTFILES \\
NAMEFILE2        & Name of the output file for the evolution of the nuclides whose & \\
                 & evolution has to be followed & \\
\hline
NBH              & Adimensional baryon number density, $n_B/m_e^3$ & OUTVAR \\
THETAH           & Adimensional Hubble function times 3, $3\hH$ & \\
TXH              & Neutrino to photon temperature ratio, $T_X/T$ & \\
\hline
DZP              & Previous iteration value of the step-size of the independent variable & PREVVAL \\
                 & in the resolution of the nucleosynthesis equations & \\
SUMMYP           & Value of the linear combination $\sum_i \left(\hdmi + \frac{3}{2\, z} \right)\, \hgi$ & \\
SUMZYP           & Value of the linear combination $\sum_i~ Z_i\, \hgi$ & \\
ZP               & Previous iteration value of the evolution variable $z=m_e/T$ & \\
\hline
CFLAG            & Flag for the input variable type in the card reading (card mode) & READINP \\
IKEY             & Progressive argument key number in the card reading (card mode) & \\
ISTART           & Starting point of the line in the card reading (card mode) & \\
DNCHRAT          & Number of reactions to be added to the changed ones & \\
LINE             & Line input from the card file (card mode) & \\
\hline
IFORM(NREC)      & Reaction type (1-12) & RECPAR \\
NG(NREC)         & Number of incoming nuclides of type TG & \\
NH(NREC)         & Number of incoming nuclides of type TH & \\
NI(NREC)         & Number of incoming nuclides of type TI & \\
NJ(NREC)         & Number of incoming nuclides of type TJ & \\
NK(NREC)         & Number of incoming nuclides of type TK & \\
NL(NREC)         & Number of incoming nuclides of type TL & \\
Q9(NREC)         & Energy released in reaction (in unit of $10^9$ K) & \\
REV(NREC)        & Reverse reaction coefficient & \\
TG(NREC)         & Incoming nuclide type & \\
TH(NREC)         & Outgoing nuclide type & \\
TI(NREC)         & Incoming nuclide type & \\
TJ(NREC)         & Incoming nuclide type & \\
TK(NREC)         & Outgoing nuclide type & \\
TL(NREC)         & Outgoing nuclide type & \\
\hline
RATEPAR(NREC,13) & Reaction parameter values (=IFORM+TI+...+NI+...) & RECPAR0 \\
\hline
RSTRING(NREC)    & Reaction text strings & RSTRINGS \\
\hline
LH               & Function $\lh$ & THERMQ \\
LHPHI            & Derivative of $\lh$ with respect to $\phi_e$ & \\
LHZ              & Derivative of $\lh$ with respect to $z$ & \\
NAUX             & Neutrino auxiliary function, $\mathcal{N}(z)$ & \\
PBH              & Adimensional baryon pressure, $\hp_B$ & \\
PEH              & Adimensional electron pressure, $\hp_e$ & \\
PGH              & Adimensional gamma pressure, $\hp_\gamma$ & \\
RHOBH            & Adimensional baryon energy density, $\hrho_B$ & \\
RHOEH            & Adimensional electron energy density, $\hrho_e$ & \\
RHOEHPHI         & Derivative of $\hrho_e$ with respect to $\phi_e$ & \\
RHOEHZ           & Derivative of $\hrho_e$ with respect to $z$ & \\
RHOGH            & Adimensional gamma energy density, $\hrho_\gamma$ & \\
RHOGHZ           & Derivative of $\hrho_\gamma$ with respect to $z$ & \\
RHOH             & Adimensional total energy density, $\hrho$ & \\
\hline
GNUC(NNUC)       & Nuclide spin degrees of freedom & SPINDF \\
\hline
\end{tabular}
\end{center}

\begin{center}
\begin{tabular}{|l|l|l|}
\hline
VARIABLE~~~~~~~~~~~~~~ & DESCRIPTION~~~~~~~~~~~~~~~~~~~~~~~~~~~~~~~~~~~~~~~~~~~~~~~~~~~~~~~~~~~~~~~~~~~~~~~~~ & COMMON~~~~~~~ \\
\hline
A(13)            & Forward weak reaction best-fit parameters (non degenerate case) & WEAKRATE \\
B(10)            & Reverse weak reaction best-fit parameters (non degenerate case) & \\
DA(12,NXIE)      & Forward weak reaction best-fit parameters (degenerate case) & \\
DB(12,NXIE)      & Reverse weak reaction best-fit parameters (degenerate case) & \\
QNP              & Forward weak reaction best-fit exponent parameter & \\
QNP1             & Forward weak reaction best-fit exponent parameter & \\
QPN              & Reverse weak reaction best-fit exponent parameter & \\
QPN1             & Reverse weak reaction best-fit exponent parameter & \\
\hline
ZZ(NNUC)         & Nuclide atomic charges, $Z_i$ & ZNUM \\
\hline
\end{tabular}
\end{center}

\end{document}